\documentclass[ 
reprint,
 amsmath,amssymb,
 aps,
 prb,
 prstab,
superscriptaddress,
longbibliography
]{revtex4-2}

\usepackage{graphicx} 
\usepackage{dcolumn}
\usepackage{bm}
\usepackage{multirow}
\usepackage{xcolor}
\usepackage{svg}
	
\usepackage[normalem]{ulem}  

\newcommand{\averaging} [1] {\left \langle  #1 \right\rangle}
\DeclareMathOperator\Imag{Im}

\definecolor{AMK}{RGB}{100,10,200}
\definecolor{COM}{RGB}{0,0,0}
\definecolor{AEF}{RGB}{159, 62, 213}
\definecolor{JHM}{RGB}{100,10,200}

\newcommand{\figwide}[0]{0.995}

\begin{document}

\title{Spontaneous and impulsive stimulated Raman scattering\\ from two-magnon modes in a cubic antiferromagnet}
\author{Anatolii E. Fedianin}
\email{Fedianin.A.E@mail.ioffe.ru}
\author{Alexandra M. Kalashnikova}
\affiliation{Ioffe Institute, 194021 St. Petersburg, Russia}
\author{Johan H. Mentink}
\affiliation{ Radboud University, Institute of Molecules and Materials,\\
Heyendaalseweg 135, 6525 AJ Nijmegen, The Netherlands}

\date{\today}

\begin{abstract}
Exchange interactions govern the ordering between microscopic spins and the highest-frequency spin excitations - magnons at the edge of the Brillouin zone.
As well known from spontaneous Raman scattering (RS) experiments in antiferromagnets, such magnons couple to light in the form of two-magnon modes - pairs of magnon with opposite wavevectors.
Experimental works on two-magnon modes driven by exchange perturbation in impulsive stimulated Raman scattering (ISRS) experiment posed a question about consistency between spin dynamics measured in the ISRS and RS experiments.
Here, based on an extended spin correlation pseudovector formalism, we derive the analytical expression for observables in both types of experiments to determine a possibly fundamental differences between the detected two-magnon spectra.
We find that in both cases the magnons from the edge of the Brillouin zone give the largest contribution to the measured spectra. 
However, there is the difference in the spectra which stems from the fact that RS probes population of a continuum of incoherent modes, while in the case of impulsively driven modes, they are coherent and their phase and amplitudes are detected.
We show that for the continuum of modes, the sensitivity to the phase results in a relative shift of the main peaks in the two spectra, and the spectrum of the ISRS is significantly broadened and extends to the range above the maximum two-magnon mode frequency.
Formally, this is manifested in the fact, that the RS is described by an imaginary part of the Green function only, while the ISRS is described by the absolute value and hence additionally carries information about the real part of the Green function.
We further derive two-magnon Raman tensor dispersion and the weighting factors, which define the features of the coupling to light of the modes from the different regions in the Brillouin zone.
For more complex antiferromagnets, this potentially can be used for selective excitation of two-magnon modes with specific wave vectors.
\end{abstract}

\maketitle

\section{Introduction}

Laser-induced control of exchange interaction is seen as a key to controlling magnetic ordering at ultrafast time scales.
Several experimental \cite{Mikhaylovskiy-NComm2015,BossiniNatComm2016,Shan-PRB2024,schoenfeld2023dynamical} and theoretical \cite{mentink2014prl,Mentink-NComm2015,fabiani2022parametrically}  works have demonstrated the feasibility of such a control in antiferromagnets, and the main fingerprint of laser-induced changes of exchange coupling appeared to be excitation of so-called two-magnon modes.
These excitations comprise coupled pairs of magnons with opposite wavevectors from the whole Brillouin Zone (BZ).
Most importantly, the dominating contribution is provided by two-magnon modes with the shortest wavevectors at the edge of BZ and the highest frequencies in THz range defined by the strength of exchange interaction.
Thus, excitation of such modes is seen as a pathway towards ultimately fast nanoscale magnonics.

At the same time, two-magnon modes pose challenges for the development of intuitive theoretical descriptions, since the dynamics stems from pairs of magnons, which by definition goes beyond the well-known single-particle description.
For instance, recently it was shown that spin correlations are well suited for such purpose \cite{fedianin2023selection, formisano2024coherent}, and the laser-driven two-magnon modes can be represented as oscillations of this parameter, rather than the standard magnetic order parameters like N\'eel vector and magnetization.

The possibility of laser-induced excitation of two-magnon modes stems from the fact that they readily couple to light via a Raman process, or inelastic scattering. 
Spontaneous Raman scattering (RS) on thermal two-magnon modes in antiferromagnets was extensively studied since the 1960th, both experimentally and theoretically \cite{loudon1968theory,elliott1969effects,balucani1973theory}.
As a result, the selection rules for Raman scattering of magnon pairs were successfully described from the symmetry of the Raman tensor. 
Moreover, this enabled understanding the nontrivial shape of the spontaneous Raman scattering spectrum.
The latter is found to be governed by the magnon density of states (DOS) and, additionally, by magnon-magnon scattering.
In particular, although two-magnon modes from the whole BZ are allowed to couple to light because the momentum conservation law is satisfied, the scattering spectrum is dominated by the modes with frequencies of van Hove singularities, which are found close to edges of the BZ \cite{loudon1968theory}.
Despite the fact that the Raman spectrum is not directly sensitive to the individual wave vectors involved, still the selection rules helped to identify specific wavevector dependent weighting factors.
These highlight the contribution of scattering of magnon pairs from a particular region in the Brillouin zone.

Excitation of two-magnon modes by impulsive nondissipative impact of short laser pulses can be described as an impulsive stimulated Raman scattering (ISRS), which, in principle, is governed by the same light-matter interaction as RS \cite{Yan-JCP1987-I,Yan-JCP1987-II}. 
Indeed, it was found that symmetry of excitation of two-magnon modes can be predicted from the Raman tensor \cite{Zhao-PRB2006,Formisano-JPCM2022,fedianin2023selection}. 
Moreover, the dominating frequency of the transient optical response appears to be fairly close to the dominating frequency of two-magnon line in a RS spectrum \cite{Zhao-PRB2006,BossiniNatComm2016}. 
Nevertheless, already since the earliest works also apparent discrepancies between the dominating two-magnon frequencies in RS and ISRS were found \cite{Zhao-PRB2006}. 
These may be related to the fact that for the comparative analysis of RS and ISRS for single-particle excitations, such as optical phonons \cite{Yan-JCP1987-I,Yan-JCP1987-II}, only a small volume of the BZ near its center is relevant, whereas the excitation of RS and ISRS of the two-magnon modes govern a whole continuum of two-particle excitations over the whole BZ. 
 
Moreover, since ISRS is sensitive to both the phase and the amplitude of the detected optical response, signatures of relative dephasing may be present in the ISRS response that are absent in the RS spectrum, which only measures the occupations of the modes averaged over time under continuous driving. 
In addition, the DOS-based argument used for explaining the RS spectrum, strictly does not apply for the impulsive case, since the system is observed only transiently. 
Hence, the system does not have time to thermalize, and the occupation of magnon pairs may differ significantly from the equilibrium magnon DOS. 
Also, it remains unexplored if RS and ISRS tensors couple equally to magnons across the BZ, or if fundamentally different modes are seen with these techniques. Furthermore, there are technical differences in the detection schemes of both methods.

To gain insight in these issues, this article focuses on the difference between RS and ISRS of magnon pairs from a minimal theoretical perspective, in which we assume that the excitation and detection can be described using the established Raman tensors for the light-matter interaction. 
Specifically, we aim to understand (i) if both methods are equally sensitive to all parts of the BZ and (ii) what are the signatures of the sensitivity to the phase of the magnon-pair response that is present in ISRS but not in RS. 
To this end we generalize the existing theoretical framework where the spin dynamics is represented by dynamics of spin correlations. 
Within this framework we derive expressions for RS cross-section and for the spectrum of the transient ellipticity measured in ISRS experiments.
In both cases, we focus on the prototypical cubic Heisenberg antiferromagnet.
We confirm that both RS and ISRS are dominated by high-frequency magnons, which occupy the largest volume of the Brillouin zone. 
Nevertheless, we find that these spectra possess different shape, which is directly related to the fact that ISRS is also sensitive to the coherent, non-dissipative part of the magnon-pair response. 
Interestingly, since the magnon-pair spectrum stems from a sum of magnon-pairs throughout the whole Brillouin zone, the peak of the combined response is shifted as compared to that seen in RS. 
Moreover, due to the different detection schemes in RS and ISRS, we find that different parts of the BZ are detected in both methods. 
Depending on the dispersion of the magnon spectrum, this may be harnessed to increase sensitivity to specific magnon-pair contributions.

\section{Basic formalism}

\subsection{Cubic antiferromagnet}

\begin{figure}[t!]
\centering
\includegraphics[width=\figwide\linewidth]{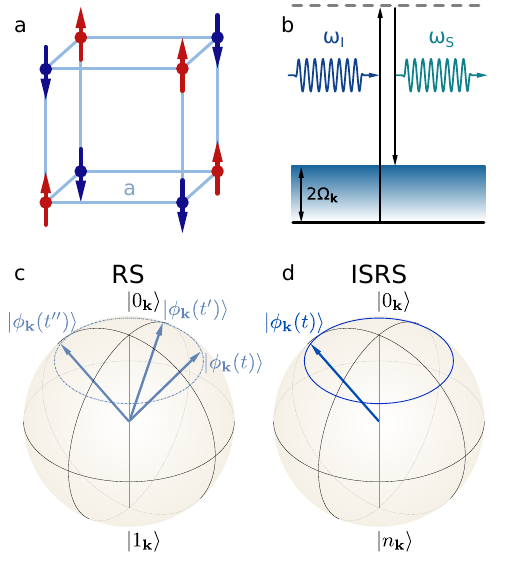}\caption{
    \label{fig:Structure}
    (a) Simple cubic crystal and antiferromagnetic ordering described by a Heisenberg model. 
    (b) Scheme of Raman scattering on the two-magnon mode with a frequency $2\Omega_k$.
    Gradient color represents increase of DOS towards higher frequencies.
    (c, d) Illustration of the difference between states of the system measured in RS and ISRS experiment. 
    Raman scattering involves excitation from the ground state $|0_\mathbf{k}\rangle$ to the excited one $|1_\mathbf{k}\rangle$.
    Due to random timing of scattering events only the population level can be measured.
    In case of ISRS, number of simultaneous scattering events is much higher and the phase of oscillations is strongly bounded with pump pulse impact moment.
    Thus, ISRS experiments contain information about oscillation's phase.
    }
\end{figure}

We consider a dielectric medium with a simple cubic crystal structure and lattice parameter $a$, and G-type antiferromagnetic ordering, as shown in Fig.~\ref{fig:Structure}(a).
The exchange interactions are described by the nearest neighbor Heisenberg model.
The magnetic unit cell of such system is doubled.
This results in a face-centered cubic (FCC) lattice with the first Brillouin zone (BZ) shown in Fig.~\ref{fig:PseudoVector}(a).
We assume that the energy band gap of the crystal is well above the laser photon energy, such that the light-matter interaction can be treated in a non-dissipative regime. 
Exemplary crystals with such parameters are KNiF$_3$ and RbMnF$_3$. 
In these materials, both RS on two-magnon modes \cite{chinn1971two,balucani1973theory} and coherent laser-induced spin dynamics at the frequency of the two-magnon mode \cite{BossiniNatComm2016,formisano2024coherent,bossini2019laser} were studied in detail. 

\subsection{Spontaneous and impulsive stimulated Raman scattering}\label{SubSec_RS_ISRS}

Spontaneous RS and ISRS are both based on an elementary inelastic scattering scheme, shown in Fig.~\ref{fig:Structure}(b).
A photon with a frequency $\omega_I$ and a wavevector $\mathbf{q}_I$ interacts non-resonantly with a medium and is scattered to a photon with $\omega_S$; $\mathbf{q}_S$ by a particular excitation in a medium. 
We are interested in a scattering by coupled pairs of magnons with frequencies $\Omega_k$ and wavevectors $\pm\mathbf{k}$ across the whole BZ.
In this case, energy and momentum conservation gives $\omega_I-\omega_S=2\Omega_k<<\omega_{I,S}$, and $\mathbf{q}_I-\mathbf{q}_S=\mathbf{k}-\mathbf{k}\approx0$.

In the case of spontaneous RS, the two-magnon pairs involved in the process are incoherent.
In RS experiments, monochromatic light of frequency $\omega_I$, polarization $\mathbf{e}_I$, and intensity $I_I(\omega_I)$ is used, and one detects an intensity of a scattered light $I_S(\omega_S)$ with polarization $\mathbf{e}_S$ as a function $\omega_S$.
The measurement relies on time-integration over many scattering events by two-magnon excitations with uncorrelated relative phases as illustrated in Fig.~\ref{fig:Structure}(c), and gives
the population of magnon pairs in the excited state. 
Thus, the continuous driving renders a steady state with an occupation of magnon pairs not present in the ground state without continuous driving. 
The scattering of light by a mode with the frequency $\omega_I-\omega_S$ is characterized by a scattering cross-section $\sigma=I_S(\omega_S)/I_I(\omega_I)$.
Conventionally, the dependence of the scattered intensity or cross-section are plotted as a function of $\omega_I-\omega_S$. 

In the case of ISRS, a system is excited by a subpicosecond laser pulse with a spectral width larger than $2\Omega_k$. 
Due to this large spectral width pairs of photons with $\omega_I-\omega_S=2\Omega_k$ within a single pulse excite two-magnon pairs with $2\Omega_k$. 
The ensemble of two-magnon pairs excited in such a way is coherent, there is a particular phase of all two-magnon pairs with a given $\mathbf{k}$, as illustrated in Fig.~\ref{fig:Structure}(d).
This makes it possible to trace the ensemble of two-magnon pairs in the time-domain.
The corresponding  modulation of the optical properties of a medium is monitored using a pump-probe technique.
While there are several realizations of optical detection \cite{Zhao-PRB2006,BossiniNatComm2016,formisano2024coherent}, 
in Ref.~\cite{fedianin2023selection} it was shown theoretically that two-magnon modes driven via ISRS give rise to transient birefringence and related oscillations of the probe pulse ellipticity $\Delta \phi'(t)$.
In order to compare the outcome of an ISRS experiment with the RS one, we use the Fourier transform of the measured signal $\Delta \phi'(t)$. 
To the best of our knowledge, only in \cite{Zhao-PRB2006,BossiniNatComm2016} the two-magnon spectra obtained in the RS and ISRS experiments were compared directly.

\subsection{Two-magnon modes and dynamics of spin correlations}

Before introducing a framework for the consistent description of two-magnon mode spectra in RS and ISRS experiments, we briefly discuss the observables used to describe antiferromagnetic system and RS and ISRS on two-magnon modes.

For the description of static and dynamic properties of magnetic media, two vectors, ferro- and antiferromagnetic (N\'eel), are conventionally used.
Strictly speaking, for antiferromagnets, these vectors do neither describe the ground state energy fully \cite{anderson1951limits}, nor spin dynamics of the two-magnon mode \cite{formisano2024coherent}.
While long-wavelength modes can be explained as the precession of macroscopic magnetization and N\'eel vectors, such description was found to fail for the short wavelength excitations as two-magnon modes at the edge of the BZ.

Spin correlations are defined as an expectation value of the scalar product of the spin operators on different positions of the crystal $\langle \hat{\mathbf{S}}_{i}(t) \cdot \hat{\mathbf{S}}_{j}(t)\rangle$.
Unlike the classical ground state of an antiferromagnet described by the N\'eel vector and corresponding spin correlations value of $-S^2$, the spin correlations in the quantum ground state can take values below $-S^2$ with a minimal value of $-S(S+1)$ corresponding to a singlet state on the given bond connecting sites denoted with $i$ and $j$. 
In the context of the Heisenberg model, the nearest neighbor spin correlation function defines the exchange energy per bond.
In terms of the spin correlations, the two-magnon mode can be seen as oscillation of the correlation function in between the classical and quantum states \cite{fedianin2023selection}.

The extended theory of RS described in classical works \cite{elliott1969effects,fleury1968scattering,loudon1968theory,cottam1986Book} relies on the evaluation of the light-matter interaction. 
For the two-magnon mode this involves the perturbation of exchange interactions and hence the Raman scattering is defined in terms of spin correlations. 
For ISRS, one can use the same light-matter interaction for describing the excitation of the spin dynamics, but assessment of the modulation of the probe pulse characteristics on the induced dynamics requires a different approach.

Initial theories of ISRS for two-magnon mode focused mainly on the modulation of the refractive index and interpreted this either as a manifestation of spin squeezing \cite{Zhao-PRL2004} or modulation of the macroscopic N\'eel vector \cite{BossiniNatComm2016,bossini2019laser}, but did not consider the relation of these observables with time-dependent spin correlations.
Importantly, the probe polarization analysis could not be resolved when considering the macroscopic N\'eel vector in the latter assessment.
Contrary, in \cite{mentink2014prl,fedianin2023selection,formisano2024coherent} nearest neighbors spin correlations $\langle \hat{\mathbf{S}}_{i}(t) \cdot \hat{\mathbf{S}}_{i+\delta}(t)\rangle$, where delta denotes the vector $\boldsymbol{\delta}$ to the nearest neighbor, were considered as a key parameter which describes spin dynamics at the frequency of the two-magnon mode.
In \cite{fedianin2023selection,formisano2024coherent} it was shown that the theory based on the equation of motion for the spin correlations themselves \textcolor{COM}{and considering only the leading order perturbation terms}, directly explains the selections rules observed in ISRS for both the pump and probe the polarization dependencies.
It was shown that the electric field of the laser pulse directed along one of the bonds perturbs the corresponding exchange coupling impulsively.
In response, periodic oscillations of spin correlations along this bond occur at the dominating frequency corresponding to the two-magnon modes at the edge of the BZ.
Furthermore, spin correlations along other bonds also oscillate with opposite phase and different amplitude, since spin correlations along different bonds share spins at particular sites.  
Such oscillations of spin correlations, in fact, bring the spin ordering along a particular bond and along other bonds closer (farther) to the N\'eel state and \textit{vice versa}, which can be seen as anisotropic change of a spin noise, or magnon squeezing, considered in \cite{Zhao-PRL2004,Zhao-PRB2006,Kamra-PRB2019}.
As a result, there are spatially anisotropic oscillations of spin correlations controlled by the polarization of the exciting pulse.
It has also been demonstrated that dynamics of spin correlations yields an anisotropic perturbation of the dielectric permittivity tensor, and, thus, result in a transient linear optical birefringence. 
Importantly, these theoretical results along with experimental verification emphasized that the antiferromagnetic vector is not a proper observable for the laser-driven two-magnon mode \textcolor{COM}{since it features negligible changes under such an excitation, given weak magnetic anisotropy of the system}  \cite{formisano2024coherent}.
Below, we use spin correlations formalism to describe both RS and ISRS experiments.

\subsection{Ground-state}

\begin{figure}
\centering
\includegraphics[width=\figwide\linewidth]{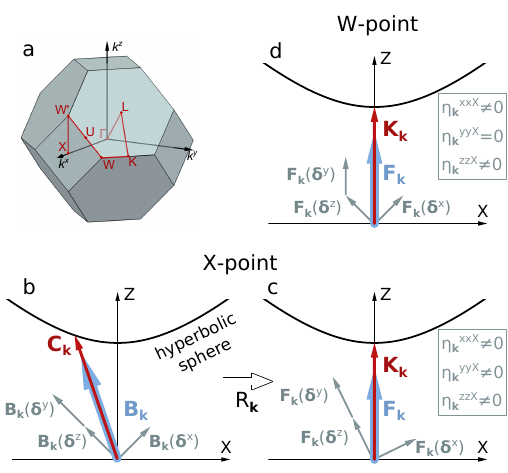}\caption{
    \label{fig:PseudoVector} 
    (a) Magnetic first BZ of the face-centered cubic antiferromagnet with the contour path and key points of BZ.
    (b) The hyperbolic sphere, the expectation value of the pseudo-vector $\langle \hat{\mathbf{C}}_\mathbf{k} \rangle$ and the effective field $\mathbf{B_k}$ in the ground state. 
    Components of $\mathbf{B}_\mathbf{k}$ produced by nearest neighbors are shown.
    The wave vector $\mathbf{k}$ is chosen at the X point of the BZ.
    (c) Rotation $R_\mathbf{k}$ of axes which transforms $\hat{\mathbf{C}}_\mathbf{k}$ into $\hat{\mathbf{K}}_\mathbf{k}$ and orients $\mathbf{B_k}$ along the $Z$ axis.
    (d) Effective exchange fields and $\langle\hat{\mathbf{K}}_\mathbf{k}\rangle$ for W point.
    In contrast to X point, $\mathbf{F}_\mathbf{k}(\boldsymbol{\delta}^y)$ is oriented along $Z$ axis, that leads to zero coupling to light polarized along $y$ axis for this $\mathbf{k}$-point.
    }
\end{figure}

We start with the Heisenberg model, which defines an unperturbed antiferromagnet with the ground state near the N\'eel state and without magnetic anisotropy. 
The Hamiltonian of such a system is
\begin{align}
    \hat{H}=J \sum_{i,\delta} \hat{\mathbf{S}}_{i} \cdot \hat{\mathbf{S}}_{i+\delta},\label{eq:Ham:Heisenberg}
\end{align}
where $J$ is an exchange energy of the system, $\hat{\mathbf{S}}_{i}=\hat{\mathbf{S}}(\mathbf{R}_i)$ is a spin operator at a position $\mathbf{R}_i$ in the lattice, the subscript $\delta$ denotes a vector $\boldsymbol{\delta}=(\delta^x,\delta^y,\delta^z)^T$ to the nearest neighbor, and $xyz$ is the real-space coordinate frame.
As noted above, the spin product terms in Eq.~\eqref{eq:Ham:Heisenberg} which enter also spin correlations, are coupled to each other since different terms contain a spin operator for the same ion.
To uncouple the terms, we proceed to $\mathbf{k}$-space and apply the bosonization operation to construct a new set of operators $\hat{\mathbf{C}}_{\mathbf{k}}$ \cite{formisano2024coherent} which linearize the Hamiltonian  [Eq.~\eqref{eq:Ham:Heisenberg}].
In this way, the spin product transforms as 
\begin{align}
    2 J \sum_i \hat{\mathbf{S}}_{i} \cdot \hat{\mathbf{S}}_{i+\delta} 
        &= -JS(S+1)N-2\sum_{\mathbf{k}} \mathbf{B}_{\mathbf{k}}(\boldsymbol{\delta}) \cdot \hat{\mathbf{C}}_{\mathbf{k}},
\end{align}
where $\hat{\mathbf{C}}_{\mathbf{k}}$ 
is the pseudovector in the hyperbolic two-magnon space:
\begin{align}
    \hat{C}^{X}_\mathbf{k}
        &=\frac{1}{2S}\left(\hat{S}^\mathcal{X}_\mathbf{k} \hat{S}^\mathcal{X}_\mathbf{-k}+\hat{S}^\mathcal{Y}_\mathbf{k} \hat{S}^\mathcal{Y}_\mathbf{-k}\right);\nonumber\\
    \hat{C}^{Y}_\mathbf{k}
        &=\frac{1}{2S}\left(\hat{S}^\mathcal{X}_\mathbf{k} \hat{S}^\mathcal{Y}_\mathbf{-k}-\hat{S}^\mathcal{Y}_\mathbf{k} \hat{S}^\mathcal{X}_\mathbf{-k}\right);\nonumber\\
    \hat{C}^{Z}_\mathbf{k}
        &=\frac{1}{2} \left(2S+1-\hat{S}^\mathcal{Z}_\mathbf{k}+ \hat{S}^\mathcal{Z}_\mathbf{-k} \right),
\end{align}
and $\mathbf{B}_{\mathbf{k}}(\boldsymbol{\delta})$ is the effective field produced by the nearest neighbor:
\begin{align}
    \mathbf{B}_{\mathbf{k}}(\boldsymbol{\delta})
        &=2 J S (-\cos(\mathbf{k}\cdot \boldsymbol{\delta}),0,1).\label{eq:BkDelta}
\end{align}
Here $N$ is the number of magnetic ions in the lattice, 
$\mathcal{XYZ}$ is the spin-space coordinate frame with $\mathcal{Z}$ being the quantization axis.
$\hat{\mathbf{C}}_{\mathbf{k}}$ has an expectation value related to the spin correlations, and satisfies the cross-product in hyperbolic space: $\hat{\mathbf{C}}_{\mathbf{k}}\times \hat{\mathbf{C}}_{\mathbf{k}}=i\hat{\mathbf{C}}_{\mathbf{k}}$.
$XYZ$ is the coordinate frame in the hyperbolic two-magnon space.
Thus, we rewrite the Hamiltonian [Eq.~\eqref{eq:Ham:Heisenberg}] as follows:
\begin{align}
    \hat{H}_{2M}&=-\sum_{\mathbf{k}} \mathbf{B}_{\mathbf{k}} \cdot \hat{\mathbf{C}}_{\mathbf{k}}, \label{eq:Ham:Ck}
\end{align}
where $\mathbf{B}_{\mathbf{k}}=\sum_{\boldsymbol{\delta}} \mathbf{B}_{\mathbf{k}}(\boldsymbol{\delta})$ is the effective field of all neighbors. 
Figures~\ref{fig:PseudoVector}(b) illustrates the relation between $\mathbf{B}_{\mathbf{k}}(\boldsymbol{\delta})$, $\mathbf{B}_{\mathbf{k}}$ and $\langle \hat{\mathbf{C}}_\mathbf{k} \rangle$ for the X point of the BZ. 

In analogy to the approach used when analyzing one-magnon excitations, 
it is convenient to align the $Z$-axis of the coordinate frame in the two-magnon hyperbolic space with the equilibrium effective field $\mathbf{B}_{\mathbf{k}}$.
For this purpose we apply an additional transformation which can be understood as the $\mathbf{k}$-dependent rotation of the system in the hyperbolic space [Fig.~\ref{fig:PseudoVector}(b,c)]: 
\begin{align}
    \hat{\mathbf{K}}_{\mathbf{k}}&=R_{\mathbf{k}} \hat{\mathbf{C}}_{\mathbf{k}}; \quad
    \mathbf{F}_{\mathbf{k}}=R_{\mathbf{k}} \mathbf{B}_{\mathbf{k}}; \nonumber\\
    R_{\mathbf{k}}
    &=\frac{1}{|\mathbf{B}_{\mathbf{k}}|}
    \begin{pmatrix}
        B^Z_{\mathbf{k}} & 0 & -B^X_{\mathbf{k}}\\
        0 & |\mathbf{B}_{\mathbf{k}}| & 0\\
        -B^X_{\mathbf{k}} & 0 & B^Z_{\mathbf{k}}
    \end{pmatrix}.
\end{align}
This rotation diagonalizes the Hamiltonian  
$\hat{H}_{2M}=\sum_{\mathbf{k}} F^{Z}_{\mathbf{k}} \hat{K}^Z_{\mathbf{k}}$, while conserving the length of the effective field, $|\mathbf{F}_{\mathbf{k}}|=|\mathbf{B}_{\mathbf{k}}|$.
In the ground state, the expectation value $\langle\hat{K}_{\mathbf{k}}^Z\rangle$ is constant and nonzero, whereas  $\langle\hat{K}^X_{\mathbf{k}}\rangle$ and $\langle\hat{K}^Y_{\mathbf{k}}\rangle$ are equal to zero.
The $\hat{\mathbf{K}}_{\mathbf{k}}$ operators introduced here are identical to the operators introduced in \cite{fabiani2022parametrically}. 

\subsection{Dielectric susceptibility and Raman tensor for two-magnon modes}

In order to describe the scattering cross-section in RS or probe polarization modulation in ISRS experiments with two-magnon modes, we introduce the dependence of dielectric susceptibility on the relevant components of $\hat{\mathbf{C}}_\mathbf{k}$ and $\hat{\mathbf{K}}_\mathbf{k}$.
In terms of the pseudo-vector $\hat{\mathbf{C}}_\mathbf{k}$ ($\hat{\mathbf{K}}_\mathbf{k}$) and the effective field $\mathbf{B}_\mathbf{k}$ $(\mathbf{F}_\mathbf{k})$, the dielectric permittivity of a cubic antiferromagnet acquires an additional diagonal contribution due to magnetic ordering 
\begin{align}
    \hat{\chi}^{\nu \nu} (\omega)
        &=-2\sum_{\mathbf{k},\boldsymbol{\delta}}\frac{\mathbf{B}_{\mathbf{k}}(\boldsymbol{\delta}) \cdot \hat{\mathbf{C}}_{\mathbf{k}} }{ U^2-\hbar^2\omega^2} \frac{Q^2 \delta^{\nu} \delta^{\nu}}{\varepsilon_0 a^{3}}\nonumber\\
         &=-2\sum_{\mathbf{k},\boldsymbol{\delta}}\frac{ \mathbf{F}_{\mathbf{k}}(\boldsymbol{\delta})\cdot \hat{\mathbf{K}}_{\mathbf{k}} }{ U^2-\hbar^2\omega^2} \frac{Q^2 \delta^{\nu} \delta^{\nu}}{\varepsilon_0 a^{3}}, \label{eq:chi1}
\end{align}
where $\nu=x,y,z$, $U$ is the energy gap of the dielectric, $Q$ is the effective charge in the Hubbard model \cite{fedianin2023selection,Kitamura2017}, and $\varepsilon_0$ is the vacuum permittivity. 
Although derived for the single-band Hubbard model, similar expressions can be obtained directly from a symmetry analysis \cite{Fleury-PRL1968}.
Expression [Eq.~\eqref{eq:chi1}] can be rewritten as follows:
\begin{align}
    \hat{\chi}^{\nu \nu} (\omega)
        &=\sum_{\mathbf{k},\Upsilon} \eta^{\nu \nu \Upsilon}_{\mathbf{k}}(\omega)\hat{K}^\Upsilon_{\mathbf{k}},\\
    \eta^{\nu \nu \Upsilon}_{\mathbf{k}}(\omega)
        &=-2\sum_{\boldsymbol{\delta}}\frac{ F^{\Upsilon}_{\mathbf{k}}(\boldsymbol{\delta})}{ U^2-\hbar^2\omega^2} \frac{Q^2 \delta^{\nu} \delta^{\nu}}{\varepsilon_0 a^{3}},
        \label{eq:RamanGeneral}
\end{align}
where $\eta^{\nu \nu\Upsilon}_{\mathbf{k}}(\omega)$ has a meaning of a two-magnon susceptibility tensor for a given
$\mathbf{k}$-point in the magnetic BZ.
The first two indices $\nu$ determine components of the dielectric permittivity in real space, and the third index $\Upsilon$ corresponds to a projection of the pseudo-vector $\hat{\mathbf{K}}$ in the two-magnon hyperbolic space. 
As can be readily seen, in the ground state magnetic ordering induces only isotropic static contribution to $\langle\hat{\chi}^{\nu\nu}\rangle=\sum_{\mathbf{k}}\eta^{\nu\nu Z}\langle\hat{K}_\mathbf{k}^Z\rangle$, 
referred to as an isotropic magnetic refraction \cite{krichevtsov1984isotropic}.
For description of RS and ISRS, we are interested in anisotropic contributions to $\langle\hat{\chi}^{\nu\nu}\rangle$ related to dynamics of $\langle\hat{\mathbf{K}}_\mathbf{k}\rangle$.

\subsection{Green function of the pseudovector $\hat{\mathbf{K}}_\mathbf{k}$ }

The dynamics of $\hat{\mathbf{K}}_{\mathbf{k}}$ is obtained by solving the equation of motion for the two-magnon mode, as described for the case of ISRS in \cite{fabiani2022parametrically,formisano2024coherent}.
However, since our goal is a comparative analysis of the RS and ISRS signals, it is more convenient to use the Green function method.
We derive an expression for the Green function of $\hat{\mathbf{K}}_\mathbf{k}$ using the diagonalized Hamiltonian for the two-magnon mode:
\begin{align}
	G(\hat{K}_{\mathbf{k}}^X, \hat{K}_{\mathbf{k}}^X|\omega')&=-\frac{\hbar}{\sqrt{2 \pi}} \frac{|\mathbf{F}_{\mathbf{k}}| \averaging{\hat{K}_{\mathbf{k}}^Z}}{\hbar^2(\omega'-i \epsilon)^2-|\mathbf{F}_{\mathbf{k}}|^2}. \label{eq:Green}
\end{align}
These Green functions have resonances at frequencies of the two-magnon modes $\omega^\prime=2\Omega_\mathbf{k}=|\mathbf{B_k}|/\hbar$.
$\epsilon$ is a phenomenological damping parameter, introduced to account for finite widths and amplitudes of these resonances.
We consider only the Green function of $\hat{K}_{\mathbf{k}}^X$, as $\hat{K}_{\mathbf{k}}^Z$ commutes with the unperturbed Hamiltonian and does not oscillate after the case of light-induced perturbation. 
We omit dynamics of $\hat{K}_{\mathbf{k}}^Y$ from further analysis since $F^{Y}_\mathbf{k}(\boldsymbol{\delta}^\nu)=0$ and a relevant susceptibility tensor component is $\eta^{\nu\nu Y}_\mathbf{k}=0$ [see Eqs.~(\ref{eq:BkDelta},\ref{eq:RamanGeneral})].
Accordingly, $\eta^{\nu\nu X}_\mathbf{k}$ has the meaning of the Raman tensor, whereas $\hat{\mathbf{K}}_{\mathbf{k}}$ is the normal coordinate \cite{fabiani2022parametrically}.

\section{Results}

\subsection{Spontaneous Raman scattering}

According to \cite{cottam1986Book}, the cross-section of the Raman scattering is defined by the average product of two dielectric susceptibilities at the frequencies of the incident and scattered light:
\begin{align}
    \sigma(\omega') &\propto e_I^\alpha e_I^\beta e_S^\mu e_S^\nu \langle \chi^{\alpha \mu} \chi^{\beta \nu}\rangle,\nonumber\\
     &\propto \sum_{\mathbf{k}, \Lambda, \Upsilon} \eta^{\alpha \alpha \Lambda}_{\mathbf{k}}(\omega_{I})\eta^{\beta \beta \Upsilon}_{\mathbf{k}}(\omega_{S})\Delta \langle\hat{K}^\Lambda_{\mathbf{k}}\hat{K}^\Upsilon_{\mathbf{k}}(\omega')\rangle
\end{align}
where $\omega'=\omega_I-\omega_S$, and the symmetry of $\hat{\eta}_\mathbf{k}$ in the cubic system is taken into account. 

We apply the fluctuation-dissipation theorem to connect the expectation value $\Delta\langle\hat{K}^\Lambda_{\mathbf{k}}\hat{K}^\Upsilon_{\mathbf{k}}(\omega')\rangle$ with the Green function $G(\hat{K}^\Lambda_{\mathbf{k}},\hat{K}^\Upsilon_{\mathbf{k}}|\omega')$: $\Delta\langle\hat{K}^\Lambda_{\mathbf{k}}\hat{K}^\Upsilon_{\mathbf{k}}(\omega')\rangle = 2\hbar [1+n_\mathrm{BE}(\omega')] \Imag G(\hat{K}^\Lambda_{\mathbf{k}},\hat{K}^\Upsilon_{\mathbf{k}}|\omega')$, where $n_\mathrm{BE}(\omega')$ is the Bose-Einstein distribution function.
As was shown above, only $\hat{K}^{X}_{\mathbf{k}}$ is relevant to RS, and we obtain for the RS cross-section:
\begin{align}
 \sigma(\omega') &\propto \mathrm{Im}\sum_{\mathbf{k}}\eta^{\alpha \alpha X}_{\mathbf{k}}(\omega_I)\eta^{\beta \beta X}_{\mathbf{k}}(\omega_S){G(\hat{K}^X_{\mathbf{k}},\hat{K}^X_{\mathbf{k}}|\omega')}.
 \label{eq:RS}
\end{align}
Thus, we obtained an expression for the spectrum of the RS cross-section of the two-magnon mode as a function of the $\hat{\mathbf{K}}_{\mathbf{k}}$ pseudovectors.
Below we assume that $\eta^{\nu\nu X}$ is the same at $\omega_I$ and $\omega_S$, as one can neglect the dispersion of dielectric permittivity when both frequencies are well below the absorption edge, and $\omega_I-\omega_S<<\omega_I$.
Note that prefactors $\eta^{\alpha \alpha X}_{\mathbf{k}}\eta^{\beta \beta X}_{\mathbf{k}}$ in Eq.~\eqref{eq:RS} have a meaning of wave-vector dependent weighting factors \cite{elliott1969effects}, i.e. they show how strongly a particular point in the BZ contributes to the scattering signal.
In contrast to \cite{elliott1969effects,balucani1973theory} and other subsequent works on RS, we separate the weighting factors from the Green functions. 
This allows us to analyze separately the light-matter interaction and the spin dynamics.

\subsection{Impulsive Stimulated Raman Scattering}

In the case of ISRS we are interested in the spectrum of the coherent two-magnon mode, which is obtained from the transient probe ellipticity measured in the time-domain. 
For a probe pulse propagating along the $z$ axis and polarized at an angle $\phi$ with respect to the $x$ axis, the corresponding expression is \cite{fedianin2023selection}  
\begin{align}
    \Delta \phi' (t) 
        & =   
        \frac{\omega_{pr} d\left[\Delta \langle\chi^{xx}(t)\rangle-\Delta \langle\chi^{yy}(t)\rangle \right]}{4  c \sqrt{\varepsilon^0}}  \sin(2 \phi), \label{eq:TimeDom}
\end{align}
where $\omega_{pr}$ is the central frequency of the probing pulse, $d$ is the thickness of the crystal, and $\varepsilon^0$ is the non-magnetic part of the dielectric permittivity of the cubic crystal.
Dispersion of $\hat\chi$ within the spectral width of the probe pulse is neglected.
Laser-driven dynamics of $\hat{\mathbf{K}}_\mathbf{k}$ contributes to modulation of the dielectric permittivity as
\begin{align}
    \Delta\langle\chi^{\nu \nu}(t)\rangle
        &=\sum_{\mathbf{k}} \eta^{\nu \nu X}_{\mathbf{k}}(\omega_{pr}) \Delta \langle\hat{K}^X_{\mathbf{k}}(t)\rangle.
\end{align}

Coherent two-magnon modes are driven by the perturbation of the exchange interaction \cite{mentink2014prl}, as follows from the dependence of the dielectric susceptibility on spin correlations [Eq.~\eqref{eq:chi1}]. 
The laser-induced perturbation of the Hamiltonian has the form
\begin{align}
    \Delta \hat{H}_{2M} = \frac{a^3 \varepsilon_0}{4} f(t)\sum_{\nu\Upsilon\mathbf{k}} \eta_\mathbf{k}^{\nu\nu\Upsilon}(\omega_{p}) E^\nu E^\nu \hat{K}^\Upsilon_\mathbf{k},
    \label{eq:H2MviaB}
\end{align}
where $f(t)$ is the envelope of the pump pulse in the time-domain.
Such a perturbation can be expressed as an additional term to the effective field $\Delta F^{\Upsilon}_\mathbf{k}(t)\propto f(t)\sum_{\nu} \eta_\mathbf{k}^{\nu\nu\Upsilon}(\omega_{p}) E^\nu E^\nu$. 
If $\Delta \mathbf{F}_\mathbf{k}$ is not collinear with the equilibrium $\mathbf{F_k}$, it leads to the precession of the pseudo-vector $\hat{\mathbf{K}}$ [Fig.~\ref{fig:PseudoVector}(d)]. 
Since $\mathbf{F_k}$ is chosen to be orientated along $Z$ axis, the condition to excite spin dynamics reduces to a requirement of a non-zero term
\begin{align}
    \Delta F^{X}_\mathbf{k}(t)\propto f(t)\sum_{\boldsymbol{\delta}} (\boldsymbol{E \cdot \delta})^2 F^{X}_{\mathbf{k}}(\boldsymbol{\delta}).
    \label{eq:DeltaB}
\end{align}
As readily seen from Eq.~(\ref{eq:DeltaB}), the maximum amplitude of the perturbation is achieved at pump polarization along one of the crystal axes \cite{fedianin2023selection}.

As the dynamics of $\hat{K}^X_\mathbf{k}$ is given by the convolution of the Green function with the pump pulse envelope in the time domain, in the frequency domain we obtain a product of the Green function [Eq.~\eqref{eq:Green}] and the spectral profile of the pump pulse $f(\omega')$: 
\begin{align}
    \Delta\langle\hat{K}^X_{\mathbf{k}}(\omega')\rangle &= \frac{a^3 \varepsilon_0}{4} f(\omega') G(\hat{K}^X_{\mathbf{k}}, \hat{K}_{\mathbf{k}}^X|\omega' ) \sum_{\nu} \eta_\mathbf{k}^{\nu\nu X}(\omega_{p}) E^\nu E^\nu . 
\end{align}
Noticeably, this expression contains both the real and the imaginary part of the Green function of $\hat{\mathbf{K}}_\mathbf{k}$. 
Indeed, the light-matter interaction is treated in the non-dissipative regime, but the subsequent dynamics contains both the coherent dynamics as well as the damping of the coherent dynamics.

Then the probe ellipticity spectrum is obtained as an absolute value of the Fourier transformed $\phi'(t)$:
\begin{widetext}
\begin{align}
    \Delta \phi' (\omega') 
        &=\frac{a^3 \varepsilon_0 \omega_{pr}  d \sin(2 \phi)}{8  c \sqrt{\varepsilon^{0}}} 
        E^\nu E^\nu f(\omega^\prime)\left|\sum_{\mathbf{k},\nu} [\eta^{x x X}_{\mathbf{k}}(\omega_{pr}) - \eta^{y y X}_{\mathbf{k}}(\omega_{pr})] \eta^{\nu\nu X}_{\mathbf{k}}(\omega_p) G(\hat{K}^X_{\mathbf{k}}, \hat{K}_{\mathbf{k}}^X|\omega' ) \right|.\label{eq:ISRS}
\end{align}
\end{widetext}

Eq.~\eqref{eq:RS} and Eq.~\eqref{eq:ISRS} are the main outcomes of our theoretical consideration, and allow us to highlight differences in the two-magnon spectra as obtained in RS and ISRS. 
First, the Raman tensor $\eta^{\nu\nu X}$ enters the weighting factors in the expressions for the two spectra in a different manner.
Second, the RS spectrum is given by the imaginary part of the Green function, while the ISRS spectrum depends on both the real and imaginary parts of the function.

\begin{figure}[t]
\centering
\includegraphics[width=\figwide\linewidth]{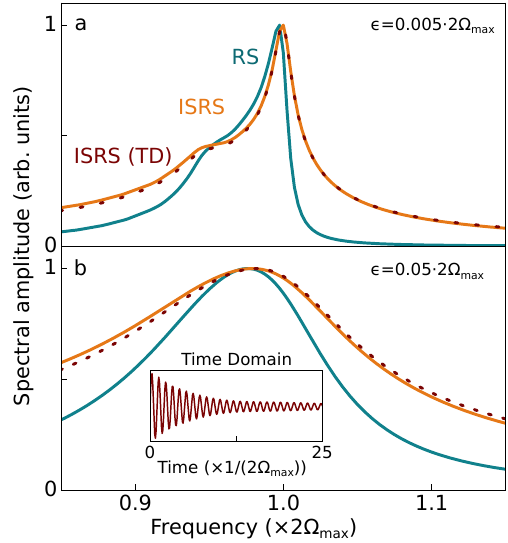}
\caption{
    \label{fig:Spectra} 
    Spectra of RS cross-section calculated using Eq.~\eqref{eq:RS} with $\alpha=\beta=x$ (light blue solid line), ISRS amplitude calculated using Eq.~\eqref{eq:ISRS} with $\nu=x$ (orange sold line), and ISRS amplitude calculated as a Fourier transform of the transient probe ellipticity [Eq.~\eqref{eq:TimeDom}] (dotted line). 
    Decay parameters are (a) $\epsilon=0.005 \cdot 2\Omega_\mathrm{max}$ and (b) $\epsilon=0.05 \cdot 2\Omega_\mathrm{max}$.
    Inset in panel (b) shows transient the probe ellipticity, calculated using Eq.~\eqref{eq:TimeDom} with $\epsilon=0.05 \cdot 2\Omega_\mathrm{max}$.
    }
\end{figure}

To illustrate the resulting difference, we consider the forward scattering geometry for RS with incident and scattered light propagating along the $z$ axis and having parallel polarizations $\mathbf{e}_I||\mathbf{e}_S||x$, giving the maximum scattering intensity.
For ISRS, we consider pump and probe pulses also propagating along the $z$ axis.
The pump is polarized along $x$ axis, while probe polarization angle is $\phi=\pi/4$, where the maximum sensitivity to the oscillations of the spin correlations is obtained \cite{formisano2024coherent}.
The magnon dispersion is shown in Fig.~\ref{fig:BZone}(a).
The spectra obtained using Eqs.~\eqref{eq:RS} and \eqref{eq:ISRS} are plotted in Fig.~\ref{fig:Spectra}(a) as a function of a normalized frequency $\omega'/2\Omega_\mathrm{max}$, where $2\Omega_\mathrm{max}=2JS/\hbar$ is the maximum frequency of the two-magnon mode which corresponds to the edge of the BZ [Fig.~\ref{fig:BZone}(a)].
The decay parameter $\epsilon=0.005 \cdot 2\Omega_\mathrm{max}$ is chosen to be low in order to emphasize special spectral points and reveal the differences between two spectra.

As expected, the dominating contributions to RS spectrum originate from two-magnon modes with $\mathbf{k}$ near the edge of BZ.
As seen in Fig.~\ref{fig:BZone}(a), the dispersion becomes flat at the edge of BZ, and there is a large density of states with the corresponding frequencies.
Applying the van Hove theorem to Eq.~\eqref{eq:RS}, we get \textcolor{COM}{two types of the singularities at the edge of the Brillouin zone. 
One of them are saddle point located at the X points, and another is the surface of maxima, which is described by $\cos(k^x a)+\cos(k^y a)+\cos(k^z a)=0$.}
This leads to emergence of a broad continuum in this spectral range with two features around $\omega'/2\Omega_\mathrm{max}=$ 1.0 and 0.95 [Fig.~\ref{fig:Spectra}(a)] corresponding to the hexagonal faces and to the centers of the square faces  of the BZ, respectively [Fig.~\ref{fig:PseudoVector}(a)].
Overall the obtained RS spectrum is in a good agreement with theoretical spectra for a cubic antiferromagnet when no magnon-magnon scattering are taken into account \cite{elliott1969effects}.

In the case of ISRS, the van Hove theorem cannot be directly applied, since one cannot replace the summation over $\mathbf{k}$ with the summation over frequencies in Eq.~\eqref{eq:ISRS}.
However, the volume in the BZ plays a dominating role in the spectrum of ISRS as well, and we find both features in the spectrum as well [Fig.~\ref{fig:Spectra}(a)].
Overall, one notices two evident differences between the RS and ISRS spectra.
First, the total ISRS spectrum is broader than the RS one. 
Second, the two local maxima, at $\omega'/2\Omega_\mathrm{max}\approx$ 1.0 and 0.95, present in both spectra, are slightly shifted with respect to each other.

\section{Discussions}

\subsection{Two-magnon Raman tensor and weighting factors dispersion} 

\begin{figure}[t]
\centering
\includegraphics[width=\figwide\linewidth]{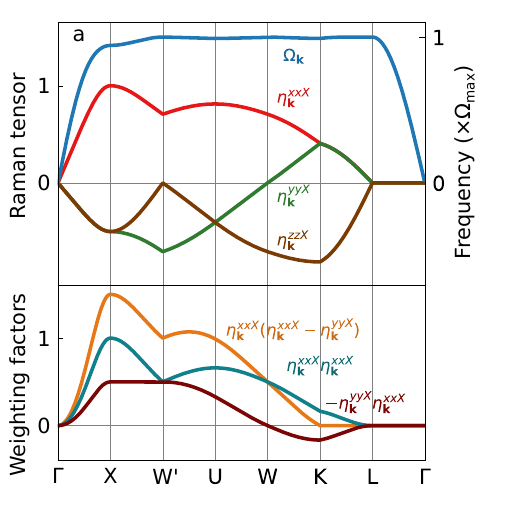}\caption{
    \label{fig:BZone} 
    (a) Magnon dispersion (blue line) and dispersion of three components of the Raman tensor $\eta^{\nu\nu X}_\mathbf{k}$, $\nu=x,y,z$ along the path in the BZ shown in Fig.~\ref{fig:PseudoVector}(a).
    (b) Dispersion of the weighting factors for RS $\eta^{xxX}_\mathbf{k}\eta^{xxX}_\mathbf{k}$ (light blue line) ISRS $\eta^{xxX}_\mathbf{k}[\eta^{xxX}_\mathbf{k}-\eta^{xxX}_\mathbf{k}]$ (orange line), and their difference $-\eta^{yyX}_\mathbf{k}\eta^{xxX}_\mathbf{k}$ (brown line).
}
\end{figure}

To understand the differences between the RS and ISRS spectra, we start from detailed examination of the Raman tensor responsible for the weighting factors entering the expressions Eq.~\eqref{eq:RS} and Eq.~\eqref{eq:ISRS}.
An explicit expression for the Raman tensor obtained from Eq.~\eqref{eq:RamanGeneral} takes the form
\begin{align}
    \eta^{\nu \nu X}_{\mathbf{k}}(\omega)
        &\propto\sum_{\boldsymbol{\delta}} F^{X}_{\mathbf{k}}(\boldsymbol{\delta})  \delta^{\nu} \delta^{\nu}\label{eq:RamanTensorExplicit}\\
        &\propto \sum_{\boldsymbol{\delta}} \frac{\frac{2}{n_a}\sum_{\boldsymbol{\delta}^\prime} \cos (\mathbf{k}\cdot \boldsymbol{\delta}^\prime) - \cos (\mathbf{k}\cdot \boldsymbol{\delta})}{|\mathbf{F}_{\mathbf{k}}|}  \delta^{\nu} \delta^{\nu},\nonumber
\end{align}
where $n_a$ stands for the number of the nearest neighbors, and is equal to 6 for the cubic lattice. 
In particular: 
\begin{align}
    \eta^{x x X}_{\mathbf{k}}(\omega)
       &\propto -\frac{1}{|\mathbf{F}_{\mathbf{k}}|} \left[2\cos (\mathbf{k}\cdot \boldsymbol{\delta}^x) -\cos (\mathbf{k}\cdot \boldsymbol{\delta}^y)-\cos (\mathbf{k}\cdot \boldsymbol{\delta}^z)\right].\label{eq:etaXX}
\end{align}
In Fig.~\ref{fig:BZone}(a) we plot the dispersion of $\eta^{\nu\nu X}_{\mathbf{k}}$, $\nu=x,y,z$ along a path in the first BZ shown in Fig.~\ref{fig:PseudoVector}(a).
Evidently, two-magnon modes from different points in BZ couple differently to light with different polarizations.
$\eta^{\nu\nu X}_\mathbf{k}=0$ at a particular point of the Brillouin zone signifies that the electric field of light with the given polarization $E^\nu$ does not couple to this particular two-magnon mode.
In terms of effective field, this corresponds to $\textbf{F}_\mathrm{k}(\delta^\nu)||Z$, i.e., its perturbation by light does not deviate the total effective field $\textbf{F}_\mathrm{k}$, as illustrated in Fig.~\ref{fig:BZone}(d) for the W point where $\eta^{yyX}=0$.

\begin{figure*}
\centering
\includegraphics[width=1\linewidth]{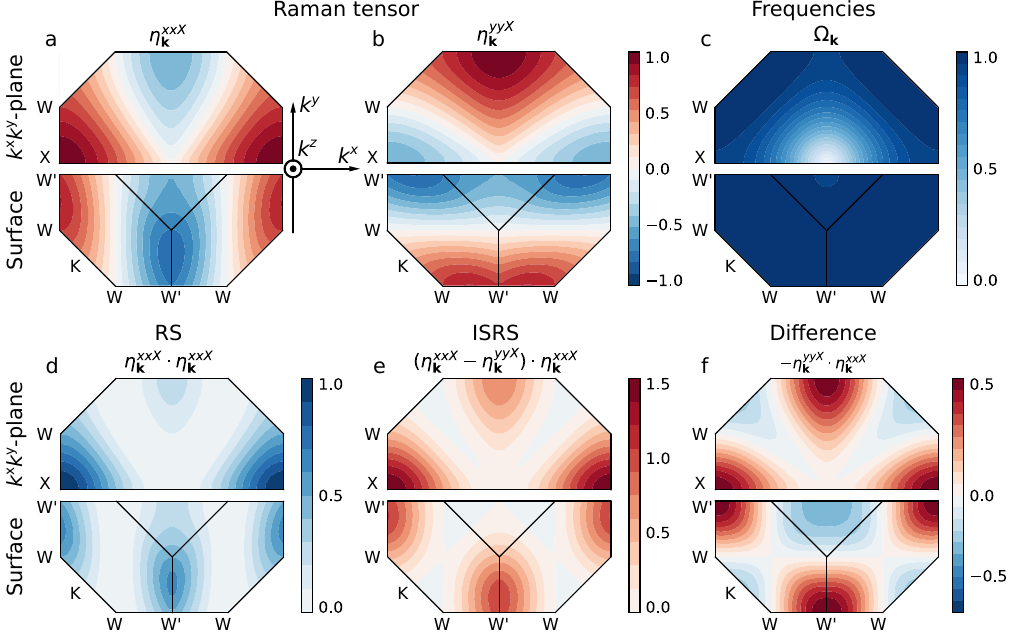}\caption{
    \label{fig:etaMap} 
    (a, b) Two-magnon Raman tensor $\eta^{\nu\nu X}_{\mathbf{k}}$, $\nu=x,y$, and (c) normalized magnon frequencies $\Omega_k/\Omega_\mathrm{max}$ in the $k^x k^y$-plane (top subpanels) and on the surface of the first BZ (bottom subpanels). 
    Projection of the surface on the $k^x k^y$ plane is shown.
    (d, e) Weighting factors for the RS and ISRS and (f) their difference in the same plane and surface.
   }
\end{figure*}

To present a more complete picture, we plot in Fig.~\ref{fig:etaMap}(a-c) how susceptibilities $\eta^{xxX}_\mathbf{k}$ and $\eta^{yyX}_\mathbf{k}$ and frequencies are distributed within the $k^x k^y$ plane and on the surface of the BZ.
These maps reveal that the absolute maxima of $\eta^{xxX}_\mathbf{k}$ and $\eta^{yyX}_\mathbf{k}$ are at the distinct X points $(\pi/a,0,0)$ and $(0,\pi/a,0)$, respectively.
The third component $\eta_\mathbf{k}^{zzX}$, by analogy with others, has a peak at the point $(0,0,\pi/a)$.
This results can be substantiated by noting that these points correspond to magnons with the wavevectors $k=\pi/a$ directed along one of the bonds.
Then, such magnon exhibits the largest susceptibility to the perturbation of the exchange coupling between the nearest neighbors along the same bond.

RS weighting factor  $\eta^{xxX}_\mathbf{k}\eta^{xxX}_\mathbf{k}$ possesses absolute maxima at the X-point $(\pi/a,0,0)$, as seen in \ref{fig:etaMap}(d).
The explicit form of the weighting factor for RS is $\left[\cos{k^x a}-1/3\left(\cos{k^x a}+\cos{k^y a}+\cos{k^z a}\right)\right]^2$, which reduces to the one reported in \cite{elliott1969effects}.
The presented data reveal that the two-magnon modes corresponding to small wavevectors weakly contribute to RS spectrum due to weak coupling to light in addition to the low volume of the corresponding BZ region.
In turn, the highest Raman tensor values in the vicinity the X point lead to the larger RS cross-section at the corresponding frequency, as compared to the one obtained based on analysis of BZ volume only yielding domination of the points belonging to the hexagonal faces of BZ.
It is easy to see also, that the RS spectra of the light polarized along $y$ axis, $\mathbf{e}_I||\mathbf{e}_S||y$, defined by the weighting factor $\eta^{yyX}_\mathbf{k}\eta^{yyX}_\mathbf{k}$ would have the same spectra as shown in Fig.~\ref{fig:Spectra}(a) because of high symmetry of the crystal, while it effectively probes a different region of the BZ.

The discussion above also holds for the ISRS spectrum, with the weighting factor $(\eta^{xxX}_\mathbf{k}-\eta^{yyX}_\mathbf{k})\eta^{xxX}$ having absolute maximum in the vicinity of X point [Figs.~\ref{fig:BZone}(b) and \ref{fig:etaMap}(e)].
Nevertheless, details of RS and ISRS prefactors dispersions differ.
To emphasize it, we plot the difference $-\eta^{yyX}_\mathbf{k}\eta^{xxX}_\mathbf{k}$ between the two prefactors in Figs.~\ref{fig:BZone}(b) and \ref{fig:etaMap}(f).
It is evident that the difference has its own dispersion.
Most noticeably, at the K-L segment presented in Fig.~\ref{fig:BZone}(b), the two-magnon modes give relatively weak but finite contribution to RS spectrum only.
The discussed difference between weighting factors in RS and ISRS stems from the fact that, by probing via transient birefringence in the ISRS experiment, one measures the phase shift between the two light waves which couple to the same two-magnon mode differently.
This either enhances or suppresses the contribution of the particular mode to the signal.  

Raman tensor and weighting factors dispersions emphasize an important feature of a coupling of two-magnon modes with light. 
For instance, from Eq.~\eqref{eq:etaXX} one readily sees that the light polarized along the $x$ axis couples to a mode formed by magnons with $\mathbf{k}\|x$ axis twice as strongly as with the magnons with $\mathbf{k}\|z$ or $y$ axis.
Thus selectivity, at least partial, in the magnon wavevector stems from the light polarization.
This adds another important difference between the two-magnon excitation by laser pulses, as compared to laser-driven one-magnon modes.
In the latter cases, selectivity regarding the wavevector of the generated or probed magnon requires spatial and spectral shaping of the laser pulses \cite{Satoh-NPhotonics2012,hortensius-NPhys2021,Filatov-APL2022,Khramova-PRB2023}.

It is important to note that one has a possibility to change the weighting factor dispersion in the expression for ISRS by choosing different probing technique. 
Indeed, in \cite{Zhao-PRL2004,Zhao-PRB2006} differential transmission was used to detect laser-driven two-magnon modes.
It is easy to show that, by using, e.g., $x-$polarized pump pulses and measuring the differential transmission of probe pulses polarized along $x$ or $y$ axis, one would detect the signal defined by the weighting factors $\eta^{xxX}\eta^{xxX}$ or $\eta^{yyX}\eta^{xxX}$, respectively.
In the former case, the dispersion of such a weighting factor corresponds to that for RS [Fig.~\ref{fig:etaMap}(d)], while in the later - to the difference [Fig.~\ref{fig:etaMap}(f)].
This shows that, in contrast to RS, ISRS provides extra flexibility in probing magnons in different parts in BZ.

In the cubic crystal, dispersion of the weighting factors results in the relative increase of the spectral amplitude in the vicinity of frequency corresponding to the X points. 
This is found to be independent on the RS or ISRS experiment, or on the detection scheme, and is related to the fact of the flat frequency dispersion with only two BZ regions having large volumes.
The apparent difference between RS and ISRS spectra [Fig.~\ref{fig:Spectra}(a)] has, thus, a different origin.

\subsection{Relative shift and broadening of RS and ISRS spectra}

Comparison of the RS and ISRS spectra [Fig.~\ref{fig:Spectra}(a)] reveal that the ISRS spectrum is considerably broader.
Furthermore, there is a shift of the features at $\omega'/2\Omega_\mathrm{max}\approx 0.94$ and 1 in the two spectra.
These differences stem from the character of the spin dynamics probed in the two cases and from the fact that a whole continuum of modes is excited. 
As detailed in Sec.~\ref{SubSec_RS_ISRS}, RS measures populations and energies of thermal two-magnon modes at equilibrium [Figs.~\ref{fig:Structure}(c)], while ISRS provides information about amplitude and phase of the coherent two-magnon modes [Fig.~\ref{fig:Structure}(d)].
We will now explain that the above mentioned spectral differences are caused by (i) sensitivity to the phase of the signal and (ii) the sensitivity to a continuum of modes.

\begin{figure}
\centering
\includegraphics[width=\figwide\linewidth]{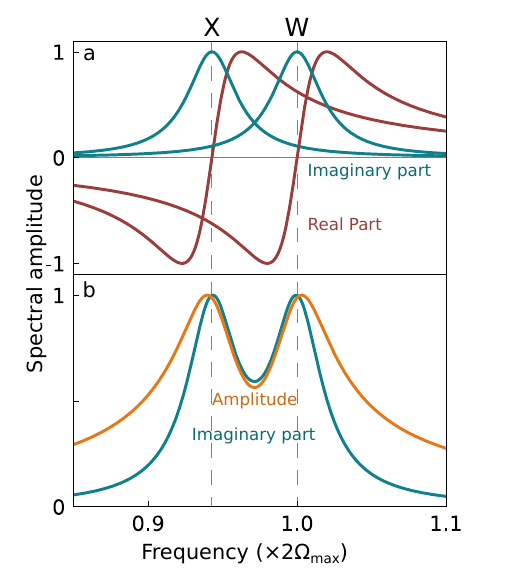}
\caption{
    \label{fig:TwoResonators}
    (a) Real (brown lines) and imaginary (light blue lines) parts spectra of two independent broad-line excitations with frequencies corresponding to the two-magnon modes at X and W points of the BZ.
    (b) Total spectrum of the imaginary part (light blue line; RS analogy) and of the amplitude (orange line; ISRS analogy) of two excitations, which shows relative shift of the peaks in the spectra and the broadening of the latter spectrum.
}
\end{figure}

To this end, we focus on a simple description of just two oscillators with finite spectral widths.
For the sake of clarity, we take oscillators with frequencies of the X and W points of the BZ [Fig.~\ref{fig:TwoResonators}(a)].
When we are interested only in the oscillators' strength, which is the case of RS, it is sufficient to consider only the imaginary parts of their response, and the resulting spectrum is given by their sum [Fig.~\ref{fig:TwoResonators}(b)].
The maxima in the spectrum are located at the eigenfrequencies of the oscillators.
When information about both the real and the imaginary parts of the spectrum is available, i.e. in the ISRS case, the amplitude spectrum can be plotted [Fig.~\ref{fig:TwoResonators}(b)].  
The two maxima in the spectrum of the oscillators appears to be shifted with respect to the eigenfrequencies of the modes, which for the case of equal oscillator strengths causes the lowest frequency peak ($X$ mode) to shift to lower frequency and the high-frequency peak ($W$ mode) to a slightly higher frequency. 
In a similar vein, for a continuum of modes we also expect shifts. 
Since the ISRS spectrum features modes with different oscillator strengths, the spectrum is asymmetric and the shift of the highest frequency peak is apparent.

Thus, we conclude that the relative shift of the spectral features in the RS and ISRS spectra should not be assigned to different modes giving leading contribution to the corresponding signals, despite the difference in the weighting factors discussed above. 
Instead, it directly follows from the fact that the two-magnon spectra in both RS and ISRS comprises contributions from many modes with different frequencies. 
Indeed, if one considers RS and ISRS spectra of one-magnon modes, only distinct modes with particular $\mathbf{k}$ contribute to the signal and their spectral lines generally do not overlap.
These are either modes with $k=0$ in a case of homogeneous laser excitation, or $k>0$ in the case of excitation using transient gratings \cite{Yan-JCP1987-I,Yan-JCP1987-II}.
As a result, although the peak can be additionally broadened due to the coherent part, no net shift is observed for the main peak \cite{Merlin-SSC1997,Imasaka-PRB2018}.

\textcolor{COM}{We note that one cannot exclude the possibility that the conventional and ultrafast infrared absorption spectroscopies of two-magnon modes would also yield different spectra.
However, our conclusions cannot be readily extended to such a case, as one needs to identify a mechanism enabling resonant interaction of light with a two-magnon mode and extend the theoretical description accordingly.}

\subsection{Comparison to experimental data}

Before proceeding to the comparison of our results with available experimental data, we note that our consideration does not include magnon-magnon scattering, \textcolor{COM}{and is applicable to the materials with negligible effect of magnon-magnon interaction.}
This effect is known to change the line shape of the RS spectrum and, in particular, to shift the high-frequency feature to lower frequencies \textcolor{COM}{since there is an additional quasi-bound state next to the non-interacting magnons \cite{Fleury-PRL1968,elliott1969effects,bouman2023time}. In general, the significance of the renormalization is limited for large $S$ and was found to be indeed weak in some antiferromagnets \cite{Prosnikov-PRB2021}.
Nevertheless, we argue that the main conclusions drawn on our analysis extend to the case with magnon-magnon interactions as well.} 
Indeed, the different way the Green function of $\hat{K}_\mathbf{k}^X$ enters the expressions for the \textcolor{COM}{RS and ISRS spectra is the same both with and without magnon-magnon interactions}. 
Specifically, taking into account magnon-magnon scattering \textcolor{COM}{leads to the renormalization of the Green function only} \cite{elliott1969effects,balucani1973theory} and does not change the Raman tensor and the weighting factors.
This, in particular, means that the relative peak shift in RS and ISRS could be still present.
We also note that, in the experimentally relevant regime of stronger decay shown in Fig.~\ref{fig:Spectra}(b), $\epsilon=0.05 \cdot 2\Omega_{max}$, where the lower-frequency feature is not pronounced, the relative shift of the main maximum is still visible.

Moreover, since in experiments the spectra are obtained using Fourier transform of the transient signals, we plot in Fig.~\ref{fig:Spectra}(a, b) the Fourier transforms of the probe ellipticity time-trace calculated using the approach described in details in \cite{formisano2024coherent,fedianin2023selection}.  
Here, exponential decay $e^{-\epsilon t}$ was introduced for each individual mode.
The time trace obtained for low value of $\epsilon$ is shown in inset in Fig.~\ref{fig:Spectra}(b).
As one can see, if the time window in time-domain measurements is sufficiently large as compared to the damping time, there is a negligible discrepancy between such spectrum and the one calculated using Eq.~\eqref{eq:ISRS}. 
Overall, apparent damping of the two-magnon signals in the time domain reported so far is sufficiently strong and is the leading effect responsible for a large spectral width of ISRS signals.

In \cite{BossiniNatComm2016,Zhao-PRB2006} two-magnon spectra obtained in the RS and ISRS experiments were directly compared.
In \cite{BossiniNatComm2016}, the cubic Heisenberg antiferromagnet KNiF$_3$ was considered, for which our model is applicable directly.
In \cite{Zhao-PRB2006} more complex tetragonal antiferromagnets FeF$_2$ and NiF$_2$ having rutile structure were studied. 
Noticeably, both works report the shift of the maxima of the ISRS spectrum with respect to the RS spectrum, in fair agreement with our theoretical analysis.
Importantly, in \cite{Zhao-PRB2006} the validity of determining the two-magnon frequency from the ISRS data was questioned.
Our analysis, indeed, confirms that one needs to keep in mind that the RS and ISRS spectra are generally different, since the coherent response of an ensemble of (nearly) degenerate modes leads to significant broadening and slight shift of the spectrum as compared to the dissipative response of the same modes.
Moreover, different states are probed due to the different dispersions in the weighting factors.
\textcolor{COM}{We note that a comparative study of RS and time-resolved RS spectra in a KNiF$_3$ was also reported in \cite{Batignani-NPhoton2015}.
However, in both these cases the RS spectrum was the observable, either in equilibrium or under the laser excitation, that is different from the detection scheme considered in the present work.}

Concomitant experimental study of the effect of Raman tensor dispersion and different weighting factors on RS and ISRS spectra is not yet available.
Such study would require materials with symmetry lower than cubic and with more features present in the magnon dispersion.
The family of transition metal fluorides, CoF$_2$, FeF$_2$, and NiF$_2$, could be promising objects for such a study.
RS in these transition metal fluorides was a subject of thorough experimental and theoretical examination \cite{Lockwood-PRB1987,Meloche-PRB2007}.
For CoF$_2$ having a strong orbital contribution to magnetic moment it was shown that two-magnon modes from different points at the edge of the BZ, namely R and M, can be distinguished in polarized RS spectra. 
It is enabled by different symmetry of the weighting factors and by measurable frequency splitting between these modes, stemming from the fact that several exchange integrals contribute to magnon frequency dispersion.
In \cite{Formisano-JPCM2022} two-magnon mode excitation was also revealed in ISRS experiment in this material.
Weakness of the reported signal, and the proximity of the strong phonon line, however, do not allow any reliable comparison of ISRS and RS spectra. 

\section{Conclusions}

In conclusion, we derived equations for spectra of RS cross-section and ISRS transient ellipticity on two-magnon mode in a single theoretical framework based on the spin correlation pseudovectors and bond-dependent exchange effective fields.  
These equations highlight the differences between the spectra which trace the differences in (i) the way light couples to the modes and (ii) the character of the spin dynamics. 
First, while RS probes the population of incoherent two-magnon modes, ISRS contains the phase information for coherent modes as well.
The RS spectrum is therefore sensitive to only the imaginary part of the Green function of the spin correlation pseudo vector, while ISRS spectrum depends on the absolute value, i.e. on both the real and imaginary part of this Green function. 
Importantly, since the two-magnon response stems from summing all modes across the BZ, differences arise that are not seen for spectrally isolated modes. 
Specifically, since the two-magnon response features spectrally overlapping modes contributions, there is a relative shift of the peaks in the two spectra, also found earlier in experiments.
Additionally, the coherent spin dynamics gives spectral weight much beyond the width determined by the damping of individual modes.

Second, we derived and analyzed two-magnon Raman tensor dispersion across BZ characterizing coupling of a magnon pair to light depending on length and direction of the magnon wavevector. 
We confirm analytically and numerically that the magnons from the edges of BZ couple to light stronger in general, owing to the nature of this coupling - light-induced perturbation of exchange having the largest impact on a short-wavelength magnons.
Furthermore, since the exchange coupling and the short-wavelength magnons are linked to crystal structure, the light polarization defines which bond is perturbed and also couples more strongly to magnon pairs with their wavevectors directed along this bound.   
The dispersion of the Raman tensor leads to the dispersion of the weighting factors fully describing the light-spin interaction in RS and ISRS experiments.
We show that the different composition of the weighting factors entering expressions for RS and ISRS spectra could, in general, yield a difference in their sensitivity to two-magnon modes in various regions of BZ. 
Furthermore, variability of the detection schemes in ISRS enables tuning these weighting factors.
In the considered cubic antiferromagnet this effect is found, however, to be not pronounced due to an almost featureless magnon frequency dispersion.
However, we believe that our finding forms a basis for further studies of ISRS in more complex antiferromagnets \cite{Lockwood-PRB1987,Meloche-PRB2007,Li-PRL2020} aimed at controllable excitation of two-magnon pairs in various regions of BZ.

\section{Acknowledgements}
We thank R.M. Dubrovin, A.V. Kimel, R.V. Pisarev, and M.A. Prosnikov for fruitful discussions.
A.M.K. acknowledges support from RSF [grant no. 23-12-00251 (https://rscf.ru/en/project/23-12-00251/)].
J.H.M. acknowledges support by the EU, in particular by the Horizon 2020 Framework Program of the European Commission under ERC-2019-SyG no. 856538 (3D MAGiC) and the Horizon Europe project no. 101070290 (NIMFEIA), as well as funding from the VIDI project no. 223.157 (CHASEMAG) which is financed by the Dutch Research Council (NWO).

The authors declare that this work has been published as a result of peer-to-peer scientific collaboration between researchers. The provided affiliations represent the actual addresses of the authors in agreement with their digital identifier (ORCID) and cannot be considered as a formal collaboration between the aforementioned institutions.

\bibliography{bibliography}

\end{document}